# Telework During the Pandemic:

# Patterns, Challenges, and Opportunities for People with Disabilities

Mason Ameri, Ph.D., Rutgers Business School, Rutgers University, mason.ameri@rutgers.edu

Douglas Kruse, Ph.D., School of Management and Labor Relations, Rutgers University, dkruse@smlr.rutgers.edu

So Ri Park, M.A., School of Management and Labor Relations, Rutgers University, sori.park@rutgers.edu

Yana Rodgers, Ph.D., School of Management and Labor Relations, Rutgers University, yana.rodgers@rutgers.edu

Lisa Schur, Ph.D.,, J.D., Department of Labor Studies and Employment Relations, Rutgers University, lschur@smlr.rutgers.edu

Corresponding author: Douglas Kruse, School of Management and Labor Relations, Rutgers University, 94 Rockafeller Road, Piscataway, NJ, 08854, dkruse@smlr.rutgers.edu, Phone 1-908-616-7841

Acknowledgments: The authors thank Sophie Mitra, Fitore Hyseni, Christine Grant, Carlo Tramontano for helpful comments. An earlier version of this paper was presented at the 2022 annual conference of the Labor and Employment Relations Association.

Funding: This line of study was supported in part by a grant from the National Institute on Disability, Independent Living, and Rehabilitation Research (NIDILRR) for the Rehabilitation Research & Training on Employment Policy: Center for Disability-Inclusive Employment Policy Research Grant [grant number #90RTEM0006-01–00] and by the RRTC on Employer Practices Leading to Successful Employment Outcomes Among People with Disabilities, Douglas Kruse PI, Grant [grant number #90RTEM0008-01-00].

Disclaimer: The views provided herein do not necessarily reflect the official policies of NIDILRR, nor do they imply endorsement by the Federal Government.

Conflict of interest: The authors declare they have no conflicts of interest.

# Telework during the Pandemic:

# Patterns, Challenges, and Opportunities for People with Disabilities

**Abstract**

**Background:** Telework has benefits for many people with disabilities. The pandemic may create new employment opportunities for people with disabilities by increasing employer acceptance of telework, but this crucially depends on the occupational structure.

**Objective:** We compare people with and without disabilities in the expansion of telework as the pandemic began, and the evolution of telework during the pandemic.

**Methods:** We use U.S. data from the American Community Survey from 2008 to 2020 and the Current Population Survey over the May 2020 to April 2022 period. Prevalence and trends are analyzed using linear probability and multinomial logit regressions.

**Results:** While workers with disabilities were more likely than those without disabilities to telework before the pandemic, they were less likely to telework during the pandemic. The occupational distribution accounts for most of this difference. Tight labor markets, as measured by state unemployment rates, particularly favor people with disabilities obtaining telework jobs. While people with cognitive/mental health and mobility impairments were the most likely to telework during the pandemic, tight labor markets especially favored the expansion of telework for people with vision impairments and difficulty with daily activities inside the home.

**Conclusions:** Many people with disabilities benefit from working at home, and the pandemic has increased employer acceptance of telework, but the current occupational distribution limits this potential. Tighter labor markets during the recovery offer hope that employers will increasingly hire people with disabilities in both telework and non-telework jobs.

**Keywords:** Disability, employment, telework, pandemic, flexibility

**Introduction**

During the COVID-19 pandemic, tens of millions of U.S. workers lost their jobs, with disproportionately larger employment losses for people with disabilities, women, and people of color.[1-2] Job losses were even greater for people with intersecting marginalized identities (such as Black women with disabilities) compared to people with no disabilities or people who are marginalized along only one dimension.[3] For others, especially those in white-collar jobs, company lockdowns forced the adoption of telecommuting arrangements.[4] Emerging evidence suggests that worker productivity, job satisfaction, and retention have improved with work from home arrangements during the pandemic.[5]

This unprecedented growth of telework may have lasting effects on employers' acceptance of telework as an accommodation for people with disabilities.[6] Regarding employment rights of people with disabilities, "Changing the location where work is performed may fall under the ADA's reasonable accommodation requirement of modifying workplace policies…"[7] However, courts have generally held that employers are not obligated to adopt worker preferences for telework, and many employers have been resistant to it.[8-10] This issue has emerged out of both case law and employer determinations, and notably, the demographic characteristics of the people requesting the accommodations affect how both judges and employers decide whether a request is reasonable.[9]

Telework can be particularly valuable for many people with disabilities.[11-12] It provides flexibility for those with impairments that impede working in traditional office settings, such as those with conditions requiring frequent breaks from work, remaining close to medical equipment, recurring medical appointments, or dealing with unpredictable flare-ups of their conditions. While valuable for all workers, reduced commuting time and expense may be especially beneficial for people with mobility impairments who find it difficult or costly to travel outside the home. Telework can also enable job retention and may help ensure that pay levels and raises are determined more by actual performance and qualifications than by stereotypes and work cultures that have been shown to disadvantage workers with disabilities.[13] Telework can be particularly helpful for older individuals with disabilities who are seeking jobs, as employers are often unwilling to hire older people with limited mobility. Viewed through the lens of the capability approach, telework can improve the "capability sets" of older people with disabilities who are seeking employment but facing the double stigma of age and disability.[14-15]

These potential advantages must be measured against the potential disadvantages of telework, which include greater social isolation, increasingly blurred lines between work and home life, and being "out of sight, out of mind" for promotion and training opportunities. The isolation could be particularly harmful given that employment is a primary means of social integration in American society. By limiting opportunities for people with disabilities to be better integrated into the social fabric of their communities, the shift to telework could have unintended adverse consequences. However, these limitations must be balanced against the likelihood that telework will increase integration for people with disabilities who would not be employed without it.

Note that telework is not feasible for every type of job. For example, many service and blue-collar jobs must be performed in person. These occupations were especially hard-hit amid the pandemic–the sectors in which people with disabilities are disproportionately employed. Sectors in which a larger proportion of workers could not work remotely experienced larger declines in employment due to the pandemic.[16] However, during tight labor markets, employers typically lower barriers to hiring, so once states began to recover from the pandemic and experienced declining unemployment, we would expect to see an increase in employment for people with disabilities and other marginalized groups. An interesting question is how tighter labor markets affect the opportunity for telework for people with disabilities.

Little is known about the effects of telework on workers with disabilities. Some evidence indicates that accommodations, including working from home, help employees with disabilities stay attached to the labor force,[17] but home-based work does not appear to reduce disability pay disparities.[11]

To address this gap, our study examines the extent to which people with disabilities worked from home due to COVID-19, how this changed as the pandemic progressed, and the role of the occupational structure and tight labor markets. We first use annual data from the American Community Survey to examine the increase in the likelihood of primarily working from home as the pandemic began in 2020. We then use monthly data from the Current Population Survey during the pandemic to explain disability differences in pandemic-related telework and the effects of tight labor markets on the probability of pandemic-related telework and non-telework employment. Results point to both the limitations of telework for people with disabilities as well as new opportunities with tightening labor markets.

## Methods

In this study, employment and telework measures are constructed using data from the American Community Survey (ACS)[18] and the Current Population Survey (CPS),[19] both of which are conducted by the Census Bureau. Both provide data on demographic characteristics as well as measures of disability based on a six-question set asked since 2008: (1) "Is this person deaf or does he/she have serious difficulty hearing?"; (2) "Is this person blind or does he/she have serious difficulty seeing even when wearing glasses?"; (3) "Because of a physical, mental, or emotional condition, does this person have serious difficulty concentrating, remembering, or making decisions?"; (4) "Does this person have serious difficulty walking or climbing stairs?"; (5) "Does this person have difficulty dressing or bathing?"; (6) "Because of a physical, mental, or emotional condition, does this person have difficulty doing errands alone such as visiting a doctor's office or shopping?" Respondents may choose more than one category, so the categories are not mutually exclusive.

The telework measure from the ACS is based on the question asked of employed people "How did this person usually get to work last week?" Those who responded "Worked from home" are counted in our study as "primarily working from home." Our other measure of telework from the CPS data specifically captures whether COVID-19 caused the respondent to have to work from home. In May 2020, the CPS started asking a special set of five questions each month related to COVID-19, including whether the individual had worked from home for pay because of the pandemic in the past four weeks.[20] An important note is that the question specified that the work at home had to occur because of the pandemic, so pre-existing home-based work is not counted in the CPS.

The ACS data are used first to compare primarily working from home by disability status, focusing on 2019 and 2020 as the pandemic began using a simple "difference-in-difference" type of approach. Technically, we do not have a treated and control group since both people with and without disabilities experienced changes in telework patterns due to the pandemic. However, the approach works well to describe how telework patterns pre- and post-pandemic differ between people with and without disabilities. In this framework, whether or not individual *i* is engaged in telework (T) in year *t* is expressed as follows:

$$T_{it} = \alpha + \beta_1 D_{it} + \beta_2 Y_{it} + \beta_3 D_{it} * Y_{it} + \beta_4 X_{it} + e_{it}, \qquad (1)$$

The notation D denotes whether or not someone has a disability, Y denotes a dummy variable for the year 2020 when the pandemic started, and X is a vector of control variables. The coefficient $\beta_3$ on the interaction between disability status and the year 2020 is the key parameter of interest and captures the estimate of the differential effect of COVID-19 by disability status.

We then use CPS data to concentrate on telework during the pandemic. Missing data are imputed by the Census Bureau using standard algorithms. We use the CPS data from May 2020 to April 2022 to run linear probability regressions to predict the likelihood of telework among all employed people, first with disability status only, then including detailed occupations, and then including detailed industries and demographic characteristics (gender, age, race/ethnicity, education, number of children under age 18, part-time versus full-time status, and employee versus self-employed status). We then repeat these regressions using the six measures of disability, and the number of disabling conditions, as predictors.

Finally, we use the CPS data from May 2020 to April 2022 for the entire adult population (both employed and non-employed) to estimate the effect of tight labor markets on telework during the pandemic, using multinomial logit regressions that show the effect of a one-point decline in the state unemployment rate on the probability of employment. These effects are estimated separately for people with and without disabilities, and by disability type and number of disabling conditions. We include controls for gender, age, race/ethnicity, month, state, and state interacted with disability. Month fixed effects are included for every month in the period, to account for general trends over the pandemic. The state fixed effects are important for capturing time-invariant state-specific characteristics, and the interactions between states and disability status capture time-invariant state-specific factors related to disability, such as employment laws and institutional and geographical characteristics that affect employment of people with disabilities. We probe the results by restricting to working-aged people, and by including occupational controls which restricts the sample to individuals with stronger connections to the labor market (because occupation is available only for those who have been employed within the past year).

## Results

Looking first at the pattern since the disability questions were first asked in 2008, the percentage of people working primarily from home has slowly risen, but this trend changed abruptly in 2020 when telework surged with the onset of the pandemic (Figure 1 using ACS data). Up through 2019, more people with disabilities engaged in telework than people without disabilities. This pattern changed in 2020 when 16.0% of people without disabilities worked from home due to the pandemic compared to 14.1% of people with disabilities, indicating that workers with disabilities were left behind in the rapid expansion of telework. Hence in 2020, people with disabilities had a lower overall rate of telework compared to people without disabilities, but this is primarily because people without disabilities saw a relatively larger jump between 2019 and 2020. People with disabilities also saw a large jump in telework, but it was not as dramatic as that of people without disabilities.

The substantial increase between 2019 and 2020 in the percentage of employed people who primarily worked from home is broken out by several characteristics in Tables 1 and 2, based on regressions using 2019-2020 data in a difference-in-difference framework.

In Table 1, the disability base effect in column 1 for the overall sample shows that people with disabilities were 1.7 percentage points more likely than those without disabilities to primarily work from home in 2019. Column 2 shows a significant upward jump of 9.8 points in 2020 for people without disabilities, while the disability interaction in column 3 shows that this increase was 2.5 points smaller among people with disabilities. Descriptive statistics and full regression results are provided in Appendix Table A-1. As a robustness check on the use of a difference-in-difference type of approach to describe changes in telework patterns for people with and without disabilities, regressions accounting for prior trends using the full 2008-2020 data are also presented in Table A-1. The parallel trends assumption cannot be rejected, and the 2020 dummy and interaction coefficients are almost identical to those obtained when using just the 2019-2020 data.

The remainder of Table 1 examines these changes by disability type and demographic characteristics. Pre-pandemic telework was higher among each of the disability types, while the increase was lower for each disability type compared to those without disabilities. The lower increase for people with disabilities occurred for both women and men and across age categories. The general increase was highest for Asian individuals, White non-Hispanic people, those of other races/ethnicities, and those with Bachelor's or graduate degrees, while people with disabilities lagged in each of these groups. The higher levels and increases among those with more education reflect their greater likelihood of being in white-collar occupations amenable to telework. Those with Associate's degrees were the only group where people with disabilities had a statistically significant increase in telework that was larger than for those without disabilities, suggesting that Associate's degrees can particularly help people with disabilities obtain office jobs where telework is possible. Unsurprisingly, the increase in home-based work was much stronger among those with internet access at home.

We recognize the potential for false positives in providing a large number of subgroup analyses (we will obtain false positives about 5% of the time when using the p<.05 significance level). We are therefore cautious in this and subsequent tables about interpreting any one subgroup result, and pay more attention to the overall pattern of results.

The increase in telework in 2020 across occupations and employee status is analyzed in Table 2. As expected, people who worked in white-collar occupations reported both higher likelihoods of telework before the pandemic and larger increases as the pandemic broke out relative to those in other occupations. These results are consistent with prior findings that in the early part of the pandemic, rates of telework were substantially higher in white-collar occupations considered suitable for telework according to Occupational Information Network (O*NET) measures.[4] Relatively few people began working primarily from home in blue-collar and service occupations where workers with disabilities are concentrated. The increase among people with disabilities lagged significantly behind that of people without disabilities only among white-collar and sales occupations. Only in administrative support was the increase significantly higher for people with disabilities than those without disabilities. While this one finding of disability's positive effect in a subgroup may simply reflect a random false positive, the finding dovetails with Table 1's result of a greater increase in telework for people with disabilities who have Associate's degrees that are often required in higher-level administrative jobs.

Table 2 also shows that the increase in home-based work was higher for employees than for the self-employed, and the increase for people with disabilities lagged significantly only among employees. Self-employed people were already more likely to work at home before the pandemic and thus probably had less potential for expanded telework.

In further breakdowns reported in Appendix Table A-2, we explore the intersection of gender and disability with other categories.

We now turn from ACS annual data—comparing telework before and after the pandemic began—to CPS monthly data examining the evolution of telework over the course of the pandemic. Among employed people, Figure 2 using CPS data shows that for most of the pandemic period, more people without disabilities did pandemic-related telework compared to people with disabilities. The gap was as large as 10.3 percentage points in May 2020, when 36.1% of employed people without a disability engaged in pandemic-related telework compared to 25.8% of employed people with a disability. The likelihood of engaging in pandemic-related telework did not converge until October 2021, when 11.8% of workers with and without disabilities engaged in pandemic-related telework. There was only one month (February 2022) when a higher share of people with disabilities engaged in pandemic-related telework than people without disabilities.

We predict pandemic-related telework controlling for other characteristics in Table 3, using linear probability models applied to the CPS data (descriptive statistics are in Appendix Table A-3). Column 1 shows that people with disabilities were 2.7 percentage points less likely than people without disabilities to engage in pandemic-related telework, controlling only for the survey month. This highly significant gap decreases to a non-significant 0.3-point difference when including detailed occupation, indicating that the lower overall rate of telework among people with disabilities is primarily due to their higher likelihood of being in blue-collar and service occupations. The coefficient remains small (1.6 percentage points) but becomes positive and statistically significant in column 3 when further including detailed industry and demographic characteristics. The higher likelihood of telework among people with disabilities when including detailed characteristics is consistent with pre-pandemic data shown above and in prior studies in the U.S.[11] and U.K.[21]

Column 3 also shows that pandemic-related telework is higher among women than men and especially high among those with college and graduate degrees. Moreover, the likelihood of telework drops with age. Given that the risk of disability increases with age, it is unsurprising that adding age as a predictor affects the disability coefficient. As a robustness check, the regressions were redone using age and age-squared as continuous variables; the results in Appendix Table A-4 show that disability coefficients are almost identical to the Table 3 regressions using age categories.

Table 3 shows that the likelihood of telework is lower among both part-time and self-employed workers. The most likely explanation for the negative coefficient for self-employment is that, as suggested above concerning the ACS data, self-employed people were already more likely to work from home before the pandemic and thus had less potential for expanded telework. Also, many self-employed people operate businesses that are not amenable to telework, such as small restaurants.

Differences across disability types are analyzed in columns 4 to 6. Controlling only for survey month, column 4 shows that five of the six disability types are negatively associated with the likelihood of teleworking, and three are statistically significant. After controlling for occupation, hearing impairment is still negatively associated with the likelihood of telework, while a cognitive/mental health impairment raises the likelihood of telework. The coefficients for the other types of disabilities are small and statistically insignificant, indicating that the lower rates of pandemic-related telework for people with those disabilities are mainly due to their higher likelihood of being in occupations that are not amenable to telework.

In column 6, when we include the complete set of demographic characteristics along with occupation and industry, cognitive/mental health and mobility impairments have statistically significant positive coefficients. Earlier research indicates that work from home can especially benefit individuals with cognitive/mental health issues who may value being away from a stressful environment and need to take unscheduled breaks.[22] During the pandemic, stress at the workplace became commonplace and likely contributed to the need for people with cognitive/mental health issues to work from home. One key benefit of working from home is flexibility, which is of particular value for people who have mental impairments that make it more challenging to work in traditional workplace settings. Telework may help people with cognitive/mental health issues who have unpredictable flare-ups of their conditions that make working consistently at a job site difficult, if not impossible, and pandemic-related stress arguably could have worsened these flare-ups.[11] The pandemic also severely curtailed mobility even for people without disabilities, making it more difficult for most workers to travel to work. It is therefore not surprising that the likelihood of pandemic-related telework rose for individuals with mobility issues even after considering occupation, industry, and other characteristics.

Apart from type of disability, a greater number of disabling conditions may affect the attraction of telework. This is explored in columns 7 to 9 of Table 3. The pattern of results as controls are added is similar across those with one, two, or three or more reported conditions. Column 9 indicates that the number of disabling conditions makes a difference, as having one condition has less than half the effect (1.1 percentage points) of having two, or three or more, conditions (3.0 and 2.7 points respectively).

Do tight labor markets make a difference? Table 4 summarizes results from multinomial logit regressions showing the effects of the state unemployment rate on total employment, telework employment, and non-telework employment. This table reports marginal effects from multinomial logits showing the probability effect of a one-point decline in the state unemployment rate. Full regression results are in Appendix Table A-6, with descriptive statistics in Appendix Table A-5.

Apart from the issue of telework, these results indicate that tight labor markets particularly benefit the employment of people with disabilities. Table 4 shows that a one percentage point decline in the state unemployment rate is linked to a 0.91 point increase in employment for people without disabilities and a 0.59 point increase for people with disabilities. While the point increase is smaller for people with disabilities, it is larger relative to their low overall employment levels. The 0.59 point increase is approximately a 2.0% increase over the 30% employment rate for people with disabilities during the pandemic. In comparison, the 0.91 point increase for people without disabilities is approximately a 1.2% increase over their 73% employment rate. Tighter labor markets, therefore, especially benefit people with disabilities when measured as a proportion of existing employment.

Table 4 also indicates that tight labor markets favor people with disabilities getting telework jobs. As shown in column 3, just over half (51.8%) of the increase in employment for people with disabilities as the state's unemployment rate declined by one point is due to the rise in telework employment. In comparison, for people without disabilities, just under one-third of this increase (30.5%) is due to telework employment.

Table 4 examines the effects of tight labor markets more closely by breaking the analysis down by type of disability. The biggest effects of tight labor markets for telework employment occur for people with a vision impairment and people with difficulty with self-care inside the home. For these two groups, a one-point decline in a state's unemployment rate is associated with 1.1-point and 1.2-point increases respectively in any type of employment. Among people with vision impairments, over two-thirds of this increase (70.8%) is an increase in telework, and among people who have difficulty with self-care, almost all of this increase (92.8%) is an increase in telework.

To probe these results, in further tests we restricted the sample to working-aged people, and added occupation as a predictor. As reported in Appendix Table A-7, the results are very similar when the sample is restricted to working-aged people. When we add occupational controls, the favorable effect of tight labor markets for people with disabilities is even more apparent: non-telework employment actually declines (-0.44 points) but is more than counterbalanced by an increase in telework employment (0.79 points), resulting in a net increase in employment. For people without disabilities, the increases in telework and non-telework employment are very similar (0.25 points for each).

We also probed the results by measuring disability as number of disabling conditions in Appendix Table A-9, and found the strongest effects for those reporting only one disabling condition, indicating that tight labor markets do not particularly favor telework for people with multiple disabilities.

**Discussion**

Many workers with disabilities have conditions that make working on site during the pandemic difficult and risky. Results from this analysis indicate that people with disabilities in the U.S. had a smaller increase in telework due to the pandemic than those without disabilities and were less likely to work from home in the initial stages of the pandemic. Ironically, this outcome is the opposite of pre-pandemic patterns, when workers with disabilities were more likely to work from home.[11] This pattern of the reverse incidence of telework before and during the pandemic between people with and without disabilities is consistent with findings from the U.K.[21]

The regression results indicate that the lower overall rate of telework among people with disabilities during the pandemic is primarily due to their greater likelihood of working in blue-collar and service occupations that are difficult or impossible to perform remotely. Overall, more than half of the gap in telework between people with and without disabilities is explained by differences in occupational distribution. People with cognitive/mental health and mobility disabilities were especially likely to engage in telework during the pandemic. This finding can perhaps be explained by aspects of telework, such as less stress and commuting time, that are particularly important to people with mental illnesses or mobility impairments. The ability to have greater social distancing is another aspect of telework that is particularly important for those who have compromised immune systems.

Results from multinomial logit regressions show that tight labor markets favor people with disabilities getting telework jobs. This finding is particularly strong for people with vision impairments and people who have difficulty with self-care in the home. As states recover from the pandemic's economic crisis, tighter labor markets offer new hope that more employers will consider hiring people with disabilities in telework arrangements. We interpret this result as an increased willingness of employers to hire teleworkers during tighter labor markets, and persons with disabilities are helping to meet the higher demand. An interesting question is whether this shift in telework opportunities for persons with disabilities is temporary or more sustainable. Evidence on disability and the business cycle is scant but suggests that when labor markets are less tight (during recessions), the overall employment of people with disabilities declines.[23]

## Conclusion

These results highlight the importance of longer-term structural changes to the occupational distribution to ensure that people with disabilities are less concentrated in blue-collar and service jobs that tend to pay less and provide less job security. The pandemic has caused employers to rethink how essential job tasks can be done, which may make them more open to accommodations in general, including working from home. However, to the extent that people with disabilities are clustered in jobs that are less amenable to telework, it will be more difficult to require that employers provide this accommodation under the ADA. At the same time, even if they are allowed to work from home, it is crucial that teleworkers do not find themselves "out of sight, out of mind" and that they receive fair pay and equal opportunities for promotions.

The finding that tight labor markets especially help people with disabilities may have policy implications. The Vocational Rehabilitation system and a wide variety of federal and state policies are designed to help workers with disabilities obtain or retain employment. To some extent a tight labor market may take the place of supportive policies in increasing employment, but it may also be useful for the VR system and other programs to direct more funding to areas where job opportunities abound in order to ensure that workers with disabilities can take advantage of these opportunities.

Given the paucity of research in this area, our findings contribute to knowledge about telework among workers with disabilities. That said, our study has several limitations. First, at the time of writing, the Census Bureau had still not released the 2021 American Community Survey data, thus limiting our analysis of changes in whether or not someone primarily works from home to just the first nine months of the pandemic. Second, the disabled population may be changing as a result of COVID-19, either through Long COVID or other disabling repercussions of the public health emergency. It could be that the influx of newly disabled persons were concentrated in particular occupational sectors, work backgrounds, or demographic groups, and these changes might confound our conclusions about the role of occupational clustering in explaining the likelihood of engaging in telework. More broadly, continuing technological developments are rapidly reshaping work and increasing the feasibility of working remotely in many occupations. This is a valuable area for further research, given telework's growth trajectory and potential benefits.

**Figure 1: Work primarily at home by disability status, 2008-2020**

This figure presents trends in percent of workers working primarily at home over the 2008-2020 period by disability status, using annual American Community Survey data.

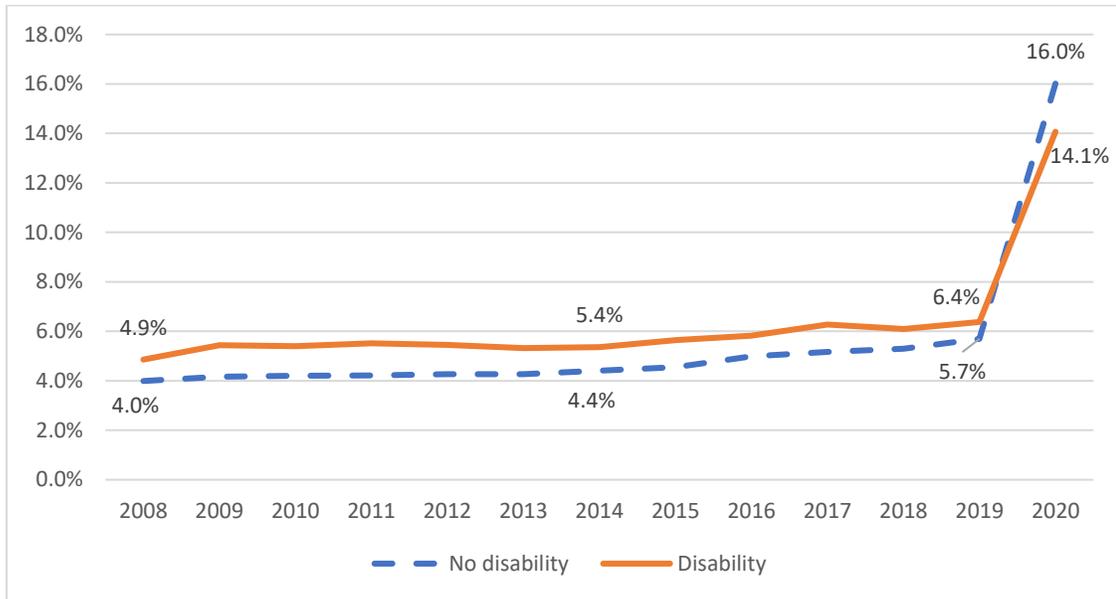

**Figure 2. Percentage of Workers Teleworking Specifically Due to the Pandemic, by Disability Status**

This figure presents trends in percent of workers working at home due to the pandemic over the May 2020 to April 2022 period by disability status, using monthly Current Population Survey data.

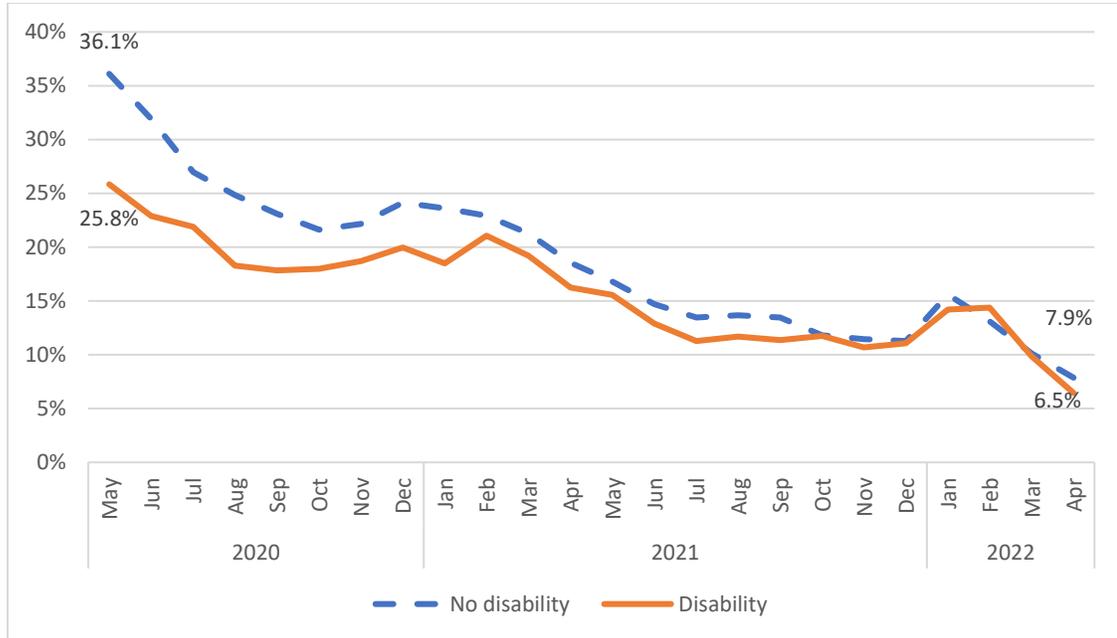

**Table 1.** Increase in Working Primarily at Home from 2019 to 2020

Each row represents separate linear probability regression, with working primarily at home as dependent variable

| | Disability base effect (2019 levels) (1) | | 2020 year dummy (2) | | Disability*2020 year dummy (3) | | N (4) | R-sq. (5) |
|---|---|---|---|---|---|---|---|---|
| Overall | 0.017** | (0.001) | 0.098** | (0.001) | -0.025** | (0.002) | 2,601,170 | 0.106 |
| **By disability type (base=no disability)** | | | | | | | | |
| Hearing impairment | 0.010** | (0.002) | 0.098** | (0.001) | -0.032** | (0.003) | 2,494,125 | 0.107 |
| Vision impairment | 0.010** | (0.002) | 0.098** | (0.001) | -0.027** | (0.004) | 2,473,758 | 0.107 |
| Cognitive impairment | 0.023** | (0.002) | 0.098** | (0.001) | -0.022** | (0.004) | 2,486,063 | 0.107 |
| Mobility impairment | 0.028** | (0.002) | 0.098** | (0.001) | -0.030** | (0.004) | 2,491,931 | 0.106 |
| Difficulty inside home | 0.034** | (0.004) | 0.098** | (0.001) | -0.019* | (0.009) | 2,450,572 | 0.107 |
| Difficulty outside home | 0.040** | (0.003) | 0.098** | (0.001) | -0.017** | (0.005) | 2,465,529 | 0.107 |
| **By gender** | | | | | | | | |
| Female | 0.023** | (0.002) | 0.108** | (0.001) | -0.027** | (0.003) | 1,245,600 | 0.107 |
| Male | 0.013** | (0.001) | 0.089** | (0.001) | -0.024** | (0.003) | 1,355,570 | 0.108 |
| **By race/ethnicity** | | | | | | | | |
| White non-Hispanic | 0.020** | (0.001) | 0.105** | (0.001) | -0.030** | (0.003) | 1,780,532 | 0.111 |
| Black non-Hispanic | 0.008** | (0.003) | 0.078** | (0.002) | -0.004 | (0.007) | 203,217 | 0.083 |
| Hispanic | 0.007* | (0.003) | 0.061** | (0.001) | -0.004 | (0.005) | 362,834 | 0.071 |
| Native American/Pacific Islander | 0.021 | (0.012) | 0.055** | (0.005) | -0.016 | (0.018) | 21,227 | 0.051 |
| Asian | 0.045** | (0.007) | 0.170** | (0.002) | -0.062** | (0.013) | 159,833 | 0.132 |
| Other race/ethnicity | 0.014* | (0.007) | 0.108** | (0.003) | -0.025* | (0.011) | 73,527 | 0.105 |
| **By age** | | | | | | | | |
| age 18-34 | 0.013** | (0.002) | 0.096** | (0.001) | -0.018** | (0.004) | 790,706 | 0.102 |
| age 35-49 | 0.018** | (0.002) | 0.109** | (0.001) | -0.027** | (0.005) | 787,327 | 0.111 |
| age 50-64 | 0.014** | (0.002) | 0.091** | (0.001) | -0.023** | (0.003) | 810,139 | 0.099 |
| age 65+ | 0.019** | (0.003) | 0.076** | (0.002) | -0.015** | (0.006) | 212,998 | 0.106 |
| **By education** | | | | | | | | |
| No HS degree | 0.004 | (0.003) | 0.021** | (0.001) | 0.002 | (0.005) | 199,061 | 0.041 |
| HS | 0.001 | (0.002) | 0.033** | (0.001) | 0.002 | (0.003) | 601,987 | 0.053 |
| Some college, no degree | 0.011** | (0.002) | 0.060** | (0.001) | 0.001 | (0.004) | 530,474 | 0.076 |
| Associate degree | 0.000 | (0.003) | 0.068** | (0.002) | 0.015* | (0.007) | 247,839 | 0.078 |
| Bachelor degree | 0.014** | (0.003) | 0.167** | (0.001) | -0.027** | (0.006) | 623,443 | 0.113 |
| Grad degree | 0.029** | (0.004) | 0.207** | (0.002) | -0.051** | (0.008) | 398,366 | 0.124 |
| **By internet access at home** | | | | | | | | |
| No internet access | 0.004 | (0.004) | 0.026** | (0.002) | 0.001 | (0.007) | 98,122 | 0.057 |
| Internet access | 0.017** | (0.001) | 0.101** | (0.001) | -0.024** | (0.002) | 2,503,048 | 0.107 |

Source: Authors' computations using American Community Survey data for 2019 and 2020.

* p<.05  ** p<.01  (standard errors in parentheses)

Control variables include dummies for gender, race/ethnicity (5), education (5), age group (3), marital status (3), any household members under age 18, any household members age 65 or older, occupation (19), employee status, and home internet access. See Table A-1 for full regression results and descriptive statistics for regression in row 1.

**Table 2.** Working Primarily at Home from 2019 to 2020, by Occupation and Employee Status

Each row represents separate linear probability regression, with working primarily at home as dependent variable

|  | Disability base effect (2019 levels) (1) | | 2020 year dummy (2) | | Disability*2020 year dummy (3) | | N (4) | R-sq. (5) |
|---|---|---|---|---|---|---|---|---|
| **By occupation** | | | | | | | | |
| **White-collar** | | | | | | | | |
|   Management | 0.026** | (0.005) | 0.147** | (0.002) | -0.056** | (0.008) | 309,995 | 0.072 |
|   Finance | 0.026** | (0.007) | 0.240** | (0.003) | -0.053** | (0.013) | 155,280 | 0.119 |
|   Computers | 0.032** | (0.010) | 0.329** | (0.004) | -0.054** | (0.020) | 92,284 | 0.160 |
|   Engineers | 0.019* | (0.008) | 0.188** | (0.004) | -0.036* | (0.018) | 60,442 | 0.108 |
|   Scientists | 0.024 | (0.013) | 0.185** | (0.005) | -0.015 | (0.027) | 31,504 | 0.100 |
|   Counselors | 0.025** | (0.009) | 0.142** | (0.004) | -0.037* | (0.017) | 49,234 | 0.068 |
|   Legal | 0.050** | (0.014) | 0.225** | (0.006) | -0.046 | (0.031) | 34,484 | 0.093 |
|   Education | 0.015** | (0.004) | 0.158** | (0.002) | -0.029** | (0.010) | 174,232 | 0.096 |
|   Entertainment | 0.041** | (0.012) | 0.176** | (0.005) | -0.047* | (0.021) | 53,552 | 0.147 |
|   Medical | 0.025** | (0.004) | 0.040** | (0.001) | -0.002 | (0.007) | 250,801 | 0.025 |
| **Services** | | | | | | | | |
|   Protective services | 0.001 | (0.004) | 0.037** | (0.002) | 0.002 | (0.009) | 54,698 | 0.022 |
|   Food services | 0.000 | (0.002) | 0.027** | (0.001) | 0.010 | (0.006) | 120,328 | 0.013 |
|   Cleaning services | -0.000 | (0.004) | 0.014** | (0.002) | 0.004 | (0.007) | 82,082 | 0.025 |
|   Personal care | 0.013 | (0.008) | 0.036** | (0.004) | 0.010 | (0.015) | 63,006 | 0.086 |
| **Sales** | 0.007 | (0.004) | 0.073** | (0.002) | -0.026** | (0.007) | 243,087 | 0.088 |
| **Administrative support** | 0.008** | (0.003) | 0.107** | (0.002) | 0.015* | (0.006) | 284,069 | 0.061 |
| **Blue-collar** | | | | | | | | |
|   Construction/extraction | -0.002 | (0.004) | 0.016** | (0.002) | -0.007 | (0.006) | 134,386 | 0.033 |
|   Repair | 0.006 | (0.004) | 0.024** | (0.002) | 0.006 | (0.009) | 83,048 | 0.040 |
|   Production | 0.002 | (0.002) | 0.025** | (0.001) | -0.005 | (0.006) | 141,891 | 0.067 |
|   Transportation | -0.001 | (0.002) | 0.019** | (0.001) | 0.007 | (0.005) | 182,767 | 0.023 |
| **By employee status** | | | | | | | | |
|   Employee | 0.017** | (0.001) | 0.104** | (0.001) | -0.026** | (0.002) | 2,323,654 | 0.095 |
| Self-employed | | 0.013* | (0.006) | 0.039** | (0.002) | -0.005 | (0.009) | 277,516 | 0.091 |

Source: Authors' computations using American Community Survey data for 2019 and 2020.

* p<.05  ** p<.01  (standard errors in parentheses)

Control variables include dummies for gender, race/ethnicity (5), education (5), age group (3), marital status (3), any household members under age 18, any household members age 65 or older, occupation (19), employee status, and home internet access. See Table A-1 for descriptive statistics on overall sample.

**Table 3.** Predicting Pandemic-related Telework

| | (1) | (2) | (3) | (4) | (5) | (6) | (7) | (8) | (9) |
|---|---|---|---|---|---|---|---|---|---|
| Disability | -0.027** | -0.003 | 0.015** | | | | | | |
| | (0.003) | (0.002) | (0.003) | - | - | - | | | |
| Disability type | - | - | - | | | | | | |
|   Hearing impairment | | | | -0.034** | -0.017** | 0.001 | | | |
| | | | | (0.005) | (0.003) | (0.003) | | | |
|   Visual impairment | | | | -0.024** | -0.009* | -0.003 | | | |
| | | | | (0.008) | (0.005) | (0.005) | | | |
|   Cognitive difficulty | | | | -0.002 | 0.027** | 0.034** | | | |
| | | | | (0.006) | (0.003) | (0.003) | | | |
|   Mobility difficulty | | | | -0.021** | -0.006* | 0.013** | | | |
| | | | | (0.006) | (0.003) | (0.003) | | | |
|   Difficulty dressing or bathing | | | | 0.023 | -0.003 | -0.008 | | | |
| | | | | (0.015) | (0.009) | (0.009) | | | |
|   Difficulty going outside alone | | | | -0.008 | 0.004 | 0.009 | | | |
| | | | | (0.010) | (0.006) | (0.006) | | | |
| Number of disabilities | | | | | | | | | |
|   One | | | | | | | -0.028** | -0.005** | 0.011** |
| | | | | | | | (0.004) | (0.002) | (0.002) |
|   Two | | | | | | | -0.031** | 0.008 | 0.030** |
| | | | | | | | (0.008) | (0.004) | (0.004) |
|   Three or more | | | | | | | -0.014 | 0.004 | 0.027** |
| | | | | | | | (0.012) | (0.007) | (0.007) |
| Female | - | - | 0.019** | - | - | 0.019** | | | 0.019** |
| | | | (0.001) | | | (0.001) | | | (0.001) |
| Race/ethnicity (White non-Hispanic excluded) | - | - | | - | - | | | | |
|   Black non-Hispanic | | | 0.002 | | | 0.002 | | | 0.002 |
| | | | (0.002) | | | (0.001) | | | (0.001) |
|   Hispanic/Latinx | | | 0.001 | | | 0.001 | | | 0.001 |

|  | (1) | (2) | (3) | (4) | (5) | (6) | (7) | (8) | (9) |
|---|---|---|---|---|---|---|---|---|---|
|  |  |  | (0.002) |  |  | (0.001) |  |  | (0.001) |
| Other race/ethnicity |  |  | 0.046** |  |  | 0.046** |  |  | 0.046** |
|  |  |  | (0.003) |  |  | (0.001) |  |  | (0.001) |
| Age (18-34 excluded) | - | - |  | - | - |  |  |  |  |
| Age 35-49 dummy |  |  | 0.001 |  |  | 0.001 |  |  | 0.001 |
|  |  |  | (0.002) |  |  | (0.001) |  |  | (0.001) |
| Age 49-64 dummy |  |  | -0.015** |  |  | -0.015** |  |  | -0.015** |
|  |  |  | (0.002) |  |  | (0.001) |  |  | (0.001) |
| Age 64-99 dummy |  |  | -0.034** |  |  | -0.033** |  |  | -0.034** |
|  |  |  | (0.002) |  |  | (0.001) |  |  | (0.001) |
| Education (no HS degree excl.) | - | - |  | - | - |  |  |  |  |
| High school degree/GED |  |  | -0.004** |  |  | -0.004** |  |  | -0.004** |
|  |  |  | (0.001) |  |  | (0.001) |  |  | (0.001) |
| Associate's degree or some college |  |  | 0.012** |  |  | 0.012** |  |  | 0.012** |
|  |  |  | (0.001) |  |  | (0.001) |  |  | (0.001) |
| Bachelor's degree |  |  | 0.091** |  |  | 0.091** |  |  | 0.091** |
|  |  |  | (0.002) |  |  | (0.001) |  |  | (0.001) |
| Graduate degree |  |  | 0.162** |  |  | 0.163** |  |  | 0.163** |
|  |  |  | (0.003) |  |  | (0.002) |  |  | (0.002) |
| Number of children under age 18 | - | - | -0.002** | - | - | -0.002** | - | - | -0.002** |
|  |  |  | (0.001) |  |  | (0.000) |  |  | (0.000) |
| Part-time worker | - | - | -0.031** | - | - | -0.032** | - | - | -0.032** |
|  |  |  | (0.001) |  |  | (0.001) |  |  | (0.001) |
| Self-employed | - | - | -0.045** | - | - | -0.045** | - | - | -0.045** |
|  |  |  | (0.002) |  |  | (0.001) |  |  | (0.001) |
| 13 month dummies | Yes | Yes | Yes | Yes | Yes | Yes | Yes | Yes | Yes |
| 524 occupation dummies | No | Yes | Yes | No | Yes | Yes | Yes | No | Yes |
| 51 industry dummies | No | No | Yes | No | No | Yes | Yes | No | No |

| R-squared | 0.032 | 0.234 | 0.265 | 0.032 | 0.234 | 0.265 | 0.032 | 0.234 | 0.265 |

**Source:** Authors' computations using Current Population Survey data for May 2020-April 2022.

Note: N=1,141,669 in all regressions. Dependent variable is whether or not teleworking. Based on linear probability regressions. Standard errors in parentheses. The notation ** is p<0.01, * is p<0.05. Descriptive statistics are in Table A-3.

**Table 4.** Effect of Tight Labor Markets on Telework during the Pandemic

Marginal effects from multinomial logits showing probability effect of 1-point decline in state unemployment rate. Based on all adults age 18+

|  | Probability effect | (t-stat) | | % of addl. jobs in telework |
|---|---|---|---|---|
|  | (1) | (2) | | (3) |
| **No disability** | | | | |
|    Any employment | 0.0091 | (0.0005) | ** | |
|    Telework employment | 0.0028 | (0.0003) | ** | 30.5% |
|    Non-telework employment | 0.0062 | (0.0005) | ** | |
| **Any disability** | | | | |
|    Any employment | 0.0059 | (0.0007) | ** | |
|    Telework employment | 0.0031 | (0.0003) | ** | 51.8% |
|    Non-telework employment | 0.0028 | (0.0007) | ** | |
| **Hearing impairment** | | | | |
|    Any employment | 0.0053 | (0.0012) | ** | |
|    Telework employment | 0.0016 | (0.0008) | * | 31.1% |
|    Non-telework employment | 0.0036 | (0.0013) | ** | |
| **Vision impairment** | | | | |
|    Any employment | 0.0107 | (0.0019) | ** | |
|    Telework employment | 0.0076 | (0.0014) | ** | 70.8% |
|    Non-telework employment | 0.0031 | (0.0020) | | |
| **Cognitive impairment** | | | | |
|    Any employment | 0.0115 | (0.0014) | ** | |
|    Telework employment | 0.0052 | (0.0007) | ** | 44.7% |
|    Non-telework employment | 0.0064 | (0.0014) | ** | |
| **Mobility impairment** | | | | |
|    Any employment | 0.0072 | (0.0011) | ** | |
|    Telework employment | 0.0036 | (0.0006) | ** | 49.5% |
|    Non-telework employment | 0.0036 | (0.0011) | ** | |
| **Difficulty inside home** | | | | |
|    Any employment | 0.0121 | (0.0028) | ** | |
|    Telework employment | 0.0112 | (0.0026) | ** | 92.8% |
|    Non-telework employment | 0.0009 | (0.0031) | | |
| **Difficulty outside home** | | | | |
|    Any employment | 0.0106 | (0.0017) | ** | |
|    Telework employment | 0.0031 | (0.0008) | ** | 29.0% |
|    Non-telework employment | 0.0075 | (0.0016) | ** | |
| n | 1,976,071 | | | |

**Source:** Authors' computations using CPS data for May 2020-April 2022.
\* $p<.05$  \*\* $p<.01$

Controls include dummies for gender, age, race/ethnicity, month, state, and disability\*state. Based on multinomial logit results in Table A-6 with descriptive statistics in Table A-5.

# Online Appendix

**Appendix Table A-1: Regression results and descriptive statistics for ACS data**

|  | 2019-2020 data ||||| 2008-2020 data |||||
|---|---|---|---|---|---|---|---|---|---|---|
|  | Coeffs (s.e.) for Table 1, row 1 ||| Mean (s.d.) || Coeffs (s.e.) ||| Mean (s.d.) ||
|  | (1) | (2) |  | (3) | (4) | (5) | (6) |  | (7) | (8) |
| **Disability** | 0.017 | (0.001) | ** | 0.059 | (0.236) | 0.011 | (0.002) | ** | 0.055 | (0.228) |
| **Year2020** | 0.098 | (0.001) | ** | 0.489 | (0.500) | 0.099 | (0.001) | ** | 0.079 | (0.270) |
| **Disability*year2020** | -0.025 | (0.002) | ** | 0.030 | (0.169) | -0.025 | (0.002) | ** | 0.005 | (0.069) |
| **Trend 2008-2020** | -- | -- |  | -- | -- | 0.0023 | (0.0000) | ** | 6.142 | (3.745) |
| **Disability*trend 2008-2020** | -- | -- |  | -- | -- | -0.0003 | (0.0002) |  | 0.346 | (1.686) |
| **Female** | 0.018 | (0.001) | ** | 0.474 | (0.499) | 0.013 | (0.000) | ** | 0.474 | (0.499) |
| **Race/ethnicity** |  |  |  |  |  |  |  |  |  |  |
| White non-Hispanic (excl.) | -- | -- |  | 0.612 | (0.487) | -- | -- |  | 0.648 | (0.478) |
| Black non-Hispanic | -0.012 | (0.001) | ** | 0.112 | (0.316) | -0.010 | (0.000) | ** | 0.109 | (0.312) |
| Hispanic | -0.013 | (0.001) | ** | 0.178 | (0.382) | -0.009 | (0.000) | ** | 0.161 | (0.368) |
| Native American/Pacific Islander | -0.013 | (0.002) | ** | 0.007 | (0.082) | -0.004 | (0.001) | ** | 0.007 | (0.082) |
| Asian | -0.012 | (0.001) | ** | 0.061 | (0.240) | -0.018 | (0.000) | ** | 0.055 | (0.228) |
| Other race/ethnicity | 0.003 | (0.002) |  | 0.030 | (0.171) | -0.001 | (0.001) |  | 0.020 | (0.138) |
| **Age** |  |  |  |  |  |  |  |  |  |  |
| Age 18-34 | -- | -- |  | 0.352 | 0.544 |  |  |  | 0.349 | (0.477) |
| Age 35-49 | 0.012 | (0.001) | ** | 0.316 | (0.465) | 0.010 | (0.000) | ** | 0.328 | (0.469) |
| Age 50-64 | 0.011 | (0.001) | ** | 0.270 | (0.444) | 0.012 | (0.000) | ** | 0.272 | (0.445) |
| Age 65+ | 0.025 | (0.001) | ** | 0.062 | (0.241) | 0.028 | (0.001) | ** | 0.051 | (0.220) |
| **Education** |  |  |  |  |  |  |  |  |  |  |
| No HS degree | -- | -- |  | 0.085 | (0.280) |  |  |  | 0.096 | (0.294) |
| HS | -0.002 | (0.001) | * | 0.238 | (0.426) | -0.004 | (0.000) | ** | 0.248 | (0.432) |

| | | | | | | | | | |
|---|---|---|---|---|---|---|---|---|---|
| Some college, no degree | 0.010 | (0.001) | ** | 0.213 | (0.409) | 0.004 | (0.000) | ** | 0.231 | (0.422) |
| Associate degree | 0.013 | (0.001) | ** | 0.091 | (0.288) | 0.005 | (0.000) | ** | 0.089 | (0.284) |
| Bachelor degree | 0.061 | (0.001) | ** | 0.233 | (0.423) | 0.028 | (0.000) | ** | 0.213 | (0.409) |
| Grad degree | 0.080 | (0.001) | ** | 0.140 | (0.347) | 0.032 | (0.000) | ** | 0.123 | (0.329) |
| **Marital status** | | | | | | | | | | |
| Married, spouse present (excl.) | -- | -- | | 0.520 | (0.500) | | | | 0.530 | (0.499) |
| Separated/divorced | -0.010 | (0.001) | ** | 0.124 | (0.330) | -0.006 | (0.000) | ** | 0.134 | (0.340) |
| Widowed | -0.009 | (0.002) | ** | 0.017 | (0.131) | -0.006 | (0.001) | ** | 0.018 | (0.134) |
| Never married | -0.006 | (0.001) | ** | 0.338 | (0.473) | -0.005 | (0.000) | ** | 0.319 | (0.466) |
| **Children under age 18 in HH** | 0.001 | (0.001) | * | 0.397 | (0.489) | 0.004 | (0.000) | ** | 0.411 | (0.492) |
| **Elders age 65+ in HH** | 0.002 | (0.001) | ** | 0.127 | (0.333) | 0.006 | (0.000) | ** | 0.105 | (0.307) |
| **Internet access at home** | 0.008 | (0.001) | ** | 0.958 | (0.200) | 0.002 | (0.000) | ** | 0.926 | (0.262) |
| **Occupation** | | | | | | | | | | |
| Management | -- | -- | | 0.111 | (0.314) | | | | 0.086 | (0.281) |
| Finance | 0.071 | (0.002) | ** | 0.057 | (0.233) | 0.026 | (0.001) | ** | 0.066 | (0.248) |
| Computer | 0.137 | (0.002) | ** | 0.035 | (0.184) | 0.074 | (0.001) | ** | 0.029 | (0.167) |
| Engineers | -0.003 | (0.002) | | 0.022 | (0.146) | -0.009 | (0.001) | ** | 0.019 | (0.137) |
| Scientists | -0.024 | (0.003) | ** | 0.011 | (0.106) | -0.022 | (0.001) | ** | 0.009 | (0.096) |
| Counselors | -0.049 | (0.002) | ** | 0.018 | (0.135) | -0.034 | (0.001) | ** | 0.017 | (0.130) |
| Legal | -0.005 | (0.003) | | 0.012 | (0.110) | -0.026 | (0.001) | ** | 0.012 | (0.107) |
| Education | -0.052 | (0.001) | ** | 0.062 | (0.241) | -0.041 | (0.001) | ** | 0.059 | (0.236) |
| Entertainment | 0.081 | (0.003) | ** | 0.020 | (0.140) | 0.067 | (0.001) | ** | 0.019 | (0.136) |
| Medical | -0.089 | (0.001) | ** | 0.096 | (0.294) | -0.052 | (0.000) | ** | 0.085 | (0.279) |
| Protective service | -0.081 | (0.001) | ** | 0.021 | (0.145) | -0.045 | (0.001) | ** | 0.022 | (0.146) |
| Food preparation/servi ng | -0.079 | (0.001) | ** | 0.053 | (0.224) | -0.044 | (0.000) | ** | 0.055 | (0.227) |
| Cleaning | -0.084 | (0.001) | ** | 0.035 | (0.183) | -0.053 | (0.001) | ** | 0.041 | (0.197) |
| Personal care | -0.049 | (0.002) | ** | 0.025 | (0.157) | -0.004 | (0.001) | ** | 0.034 | (0.181) |
| Sales | -0.030 | (0.001) | ** | 0.096 | (0.294) | -0.013 | (0.001) | ** | 0.106 | (0.308) |
| Office | -0.024 | (0.001) | ** | 0.108 | (0.311) | -0.018 | (0.000) | ** | 0.129 | (0.335) |
| Construction/extraction | -0.078 | (0.001) | ** | 0.056 | (0.229) | -0.050 | (0.001) | ** | 0.057 | (0.233) |
| Repair | -0.075 | (0.001) | ** | 0.031 | (0.174) | -0.041 | (0.001) | ** | 0.032 | (0.177) |

| | | | | | | | | |
|---|---|---|---|---|---|---|---|---|
| Production | -0.075 | (0.001) | ** | 0.055 | (0.227) | -0.040 | (0.000) | ** | 0.059 | (0.235) |
| Transportation | -0.078 | (0.001) | ** | 0.075 | (0.263) | -0.046 | (0.000) | ** | 0.065 | (0.247) |
| **Self-employed** | 0.135 | (0.001) | ** | 0.099 | (0.298) | 0.163 | (0.001) | ** | 0.098 | (0.297) |
| | | | | | | | | | |
| **Constant** | 0.034 | (0.002) | ** | -- | -- | 0.014 | (0.001) | ** | -- | -- |
| **R-squared** | 0.106 | | | | | 0.226 | | | |
| n | 2,601,170 | | | | | 17,868,550 | | | |

Source: Authors' computations using American Community Survey data for 2008 to 2020.

** $p<0.01$, * $p<0.05$ (standard errors in parentheses in cols. 2 and 6)

**Summary of Table A-2**: Women with disabilities were more likely than those without disabilities to telework before the pandemic across all of the race/ethnicity categories, while among men, this was true for only White and Asian men. Every group saw a significant general increase in telework in 2020, while the increase among people with disabilities was smaller in almost every group. The increase for people with disabilities particularly lagged among Asian women and men (5.6 and 6.6 points, respectively). Black, Hispanic, and Native American men had some of the lowest general increases in home-based work from 2019 to 2020, with no significant differences between people with and without disabilities. Also of note, women without children or elder dependents had the highest general increases in home-based work, while the smaller increases for people with disabilities did not vary much across gender and dependent status.

**Appendix Table A-2. Gender Intersections in Working Primarily at Home from 2019 to 2020**

Each row represents separate linear probability regression, with working primarily at home as dependent variable

|  | Disability base effect (2019 levels) (1) | | 2020 year dummy (2) | | Disability*2020 year dummy (3) | | N (4) | R-sq. (5) |
|---|---|---|---|---|---|---|---|---|
| **By gender and race/ethnicity** | | | | | | | | |
| Women | | | | | | | | |
| White non-Hispanic | 0.026** | (0.002) | 0.114** | (0.001) | -0.030** | (0.004) | 841,141 | 0.115 |
| Black non-Hispanic | 0.011** | (0.004) | 0.092** | (0.002) | -0.009 | (0.009) | 111,293 | 0.087 |
| Hispanic | 0.013** | (0.004) | 0.076** | (0.002) | -0.009 | (0.008) | 167,568 | 0.075 |
| Native American/Pacific Islander | 0.036* | (0.017) | 0.062** | (0.007) | -0.033 | (0.027) | 10,721 | 0.057 |
| Asian | 0.037** | (0.010) | 0.163** | (0.003) | -0.056** | (0.017) | 78,476 | 0.126 |
| Other race/ethnicity | 0.032** | (0.011) | 0.124** | (0.005) | -0.039* | (0.017) | 36,401 | 0.107 |
| Men | | | | | | | | |

|  |  |  |  |  |  |  |  |  |  |
|---|---|---|---|---|---|---|---|---|---|
| White non-Hispanic |  | 0.015** | (0.002) | 0.098** | (0.001) | -0.030** | (0.003) | 939,391 | 0.112 |
| Black non-Hispanic |  | 0.006 | (0.004) | 0.061** | (0.002) | 0.000 | (0.009) | 91,924 | 0.079 |
| Hispanic |  | 0.001 | (0.003) | 0.049** | (0.001) | 0.001 | (0.007) | 195,266 | 0.068 |
| Native American/Pacific Islander |  | 0.008 | (0.015) | 0.048** | (0.007) | -0.002 | (0.023) | 10,506 | 0.053 |
| Asian |  | 0.053** | (0.009) | 0.176** | (0.003) | -0.066** | (0.018) | 81,357 | 0.141 |
| Other race/ethnicity |  | -0.005 | (0.008) | 0.092** | (0.004) | -0.012 | (0.013) | 37,126 | 0.105 |
| **By marital status** |  |  |  |  |  |  |  |  |  |
| Women |  |  |  |  |  |  |  |  |  |
| Married, spouse present | No | 0.020** | (0.002) | 0.099** | (0.001) | -0.020** | (0.004) | 573,407 | 0.096 |
|  | Yes | 0.024** | (0.003) | 0.116** | (0.001) | -0.032** | (0.005) | 672,193 | 0.112 |
| Men |  |  |  |  |  |  |  |  |  |
| Married, spouse present | No | 0.012** | (0.002) | 0.075** | (0.001) | -0.014** | (0.004) | 542,647 | 0.100 |
|  | Yes | 0.012** | (0.002) | 0.101** | (0.001) | -0.031** | (0.004) | 812,923 | 0.110 |
| **By children or elders in household** |  |  |  |  |  |  |  |  |  |
| Women |  |  |  |  |  |  |  |  |  |
| Children under age 18 in HH | No | 0.026** | (0.002) | 0.114** | (0.001) | -0.027** | (0.004) | 766,532 | 0.107 |
|  | Yes | 0.020** | (0.003) | 0.099** | (0.001) | -0.031** | (0.006) | 479,068 | 0.110 |
| Elders age 65+ in HH | No | 0.022** | (0.002) | 0.110** | (0.001) | -0.025** | (0.004) | 1,061,347 | 0.110 |
|  | Yes | 0.023** | (0.004) | 0.090** | (0.002) | -0.026** | (0.008) | 184,253 | 0.095 |
| Men |  |  |  |  |  |  |  |  |  |
| Children under age 18 in HH | No | 0.015** | (0.002) | 0.090** | (0.001) | -0.024** | (0.003) | 843,680 | 0.109 |
|  | Yes | 0.008** | (0.002) | 0.088** | (0.001) | -0.024** | (0.005) | 511,890 | 0.107 |
| Elders age 65+ in HH | No | 0.011** | (0.001) | 0.092** | (0.001) | -0.024** | (0.003) | 1,176,388 | 0.109 |
|  | Yes | 0.013** | (0.003) | 0.065** | (0.002) | -0.012 | (0.006) | 179,182 | 0.101 |

Source: Authors' computations using American Community Survey data for 2019 and 2020.
\* p<.05  \*\* p<.01  (standard errors in parentheses)
Control variables include dummies for gender, race/ethnicity (5), education (5), age group (3), marital status (3), any household members under age 18, any household members age 65 or older, occupation (19), employee status, and home internet access. Descriptive statistics for overall sample in Table A-1.

**Appendix Table A-3: Descriptive statistics for Tables 3 and A-4**

|  | Mean (1) | (s.d.) (2) |
|---|---|---|
| **Telework** | 0.185 | (0.388) |
| **Disability** | 0.039 | (0.193) |
| **Disability type** | | |
| Hearing impairment | 0.013 | (0.114) |
| Visual impairment | 0.006 | (0.076) |
| Cognitive impairment | 0.012 | (0.108) |
| Mobility impairment | 0.013 | (0.113) |
| Difficulty dressing or bathing | 0.002 | (0.045) |
| Difficulty going outside alone | 0.005 | (0.071) |
| **Number of disabilities** | | |
| None | 0.961 | (0.193) |
| One | 0.031 | (0.172) |
| Two | 0.006 | (0.076) |
| Three or more | 0.003 | (0.050) |
| **Female** | 0.469 | (0.499) |
| **Race/ethnicity** | | |
| White non-Hispanic | 0.617 | (0.486) |
| Black non-Hispanic | 0.115 | (0.319) |
| Hispanic/Latinx | 0.179 | (0.384) |
| Other race/ethnicity | 0.089 | (0.285) |
| **Age** | | |
| Age 18-34 dummy | 0.336 | (0.472) |
| Age 35-49 dummy | 0.319 | (0.466) |
| Age 49-64 dummy | 0.276 | (0.447) |
| Age 64-99 dummy | 0.068 | (0.252) |
| Age | 42.790 | (14.286) |
| Age squared | 2037.178 | (1302.629) |
| **Education** | | |
|     No HS degree | 0.063 | (0.244) |
|     High school degree/GED | 0.258 | (0.438) |
|     Associate's degree or some college | 0.267 | (0.442) |
|     Bachelor's degree | 0.259 | (0.438) |
|     Graduate degree | 0.153 | (0.360) |
| **Number of children under age 18** | 0.587 | (1.004) |
| **Part-time worker** | 0.158 | (0.365) |
| **Self-employed** | 0.107 | (0.309) |
| **Observations** | 1141679 | |

Source: Authors' computations using Current Population Survey data for May 2020 to April 2022.

**Table A-4: Predicting Pandemic-related Telework with Continuous Age Variables**

Dep. Var.=telework. Based on linear probability regressions. Sample limited to employed.

| | (1) | (2) | (3) | (4) | (5) | (6) | (7) | (8) | (9) |
|---|---|---|---|---|---|---|---|---|---|
| Disability | -0.027** | -0.003 | 0.016** | | | | | | |
| | (0.003) | (0.002) | (0.002) | | | | | | |
| Disability type | | | | | | | | | |
|   Hearing impairment | | | | -0.034** | -0.017** | 0.004 | | | |
| | | | | (0.005) | (0.003) | (0.003) | | | |
|   Visual impairment | | | | -0.024** | -0.009* | -0.004 | | | |
| | | | | (0.008) | (0.005) | (0.005) | | | |
|   Cognitive impairment | | | | -0.002 | 0.027** | 0.034** | | | |
| | | | | (0.006) | (0.003) | (0.003) | | | |
|   Mobility impairment | | | | -0.021** | -0.006* | 0.014** | | | |
| | | | | (0.006) | (0.003) | (0.003) | | | |
|   Difficulty dressing or bathing | | | | 0.023 | -0.003 | -0.009 | | | |
| | | | | (0.015) | (0.009) | (0.009) | | | |
|   Difficulty going outside alone | | | | -0.008 | 0.004 | 0.010 | | | |
| | | | | (0.010) | (0.006) | (0.006) | | | |
| **Number of disabilities** | | | | | | | | | |
|   One | | | | | | | -0.028** | 0.005** | -0.012** |
| | | | | | | | (0.004) | (0.002) | (0.002) |
|   Two | | | | | | | -0.031** | 0.008 | 0.031** |
| | | | | | | | (0.008) | (0.004) | (0.004) |
|   Three or more | | | | | | | -0.014 | 0.004 | 0.029** |
| | | | | | | | (0.012) | (0.007) | (0.007) |
| Female | | | 0.018** | | | 0.018** | | | 0.018** |
| | | | (0.001) | | | (0.001) | | | (0.001) |

| | | | |
|---|---|---|---|
| Race/ethnicity (White non-Hispanic excluded) | | | |
| Black non-Hispanic | 0.002 | 0.002 | 0.002 |
| | (0.001) | (0.001) | (0.001) |
| Hispanic/Latinx | 0.001 | 0.001 | 0.001 |
| | (0.001) | (0.001) | (0.001) |
| Other race/ethnicity | 0.045** | 0.045** | 0.045** |
| | (0.001) | (0.001) | (0.001) |
| Age | 0.003** | 0.003** | 0.003** |
| | (0.000) | (0.000) | (0.000) |
| Age squared | -0.00004** | -0.00004** | -0.00004** |
| | (0.00000) | (0.00000) | (0.00000) |
| Education (no HS degree excluded) | | | |
| High school degree/GED | -0.004** | -0.004** | -0.004** |
| | (0.001) | (0.001) | (0.001) |
| Associate's degree or some college | 0.012** | 0.012** | 0.012** |
| | (0.001) | (0.001) | (0.001) |
| Bachelor's degree | 0.091** | 0.091** | 0.091** |
| | (0.001) | (0.001) | (0.001) |
| Graduate degree | 0.162** | 0.162** | 0.162** |
| | (0.002) | (0.002) | (0.002) |
| Number of children under age 18 | -0.003** | -0.003** | -0.003** |
| | (0.000) | (0.000) | (0.000) |
| Part-time worker | -0.029** | -0.029** | -0.029** |
| | (0.001) | (0.001) | (0.001) |
| Self-employed | -0.045** | -0.045** | -0.045** |
| | (0.001) | (0.001) | (0.001) |

| | | | | | | | | | |
|---|---|---|---|---|---|---|---|---|---|
| 23 month dummies | Yes | Yes | Yes | Yes | Yes | Yes | Yes | Yes | Yes |
| 524 occupation dummies | No | Yes | Yes | No | Yes | Yes | No | Yes | Yes |
| 51 industry dummies | No | No | Yes | No | No | Yes | No | No | Yes |
| | | | | | | | | | |
| R-squared | 0.032 | 0.234 | 0.265 | 0.032 | 0.234 | 0.265 | 0.032 | 0.234 | 0.265 |

Source: Authors' computations using Current Population Survey data for May 2020 to April 2022.

\*\* p<0.01, \* p<0.05 (standard errors in parentheses)

N=1,141,679 in all regressions. Descriptive statistics are in Table A-3.

**Table A-5: Descriptive statistics for Tables A-6 to A-10**

Figures represent means (standard deviations in parentheses)

| | Table A-6, Table A-8 panel A age 18+, and Table A-10 (1) | | Table A-8, panel A age 18-64 (2) | | Table A-8, panel B age 18+ (3) | | Table A-8, panel B age 18-64 (4) | |
|---|---|---|---|---|---|---|---|---|
| **Dependent variable** | | | | | | | | |
| No employment | 0.408 | (0.491) | 0.292 | (0.455) | 0.070 | (0.256) | 0.069 | (0.254) |
| Telework employment | 0.109 | (0.312) | 0.132 | (0.339) | 0.172 | (0.377) | 0.174 | (0.379) |
| Non-telework employment | 0.483 | (0.500) | 0.575 | (0.494) | 0.758 | (0.428) | 0.757 | (0.429) |
| | | | | | | | | |
| **Disability** | 0.120 | (0.325) | 0.076 | (0.265) | 0.042 | (0.200) | 0.037 | (0.188) |
| **Disability type** | | | | | | | | |
| Hearing impairment | 0.036 | (0.185) | 0.015 | (0.120) | 0.013 | (0.115) | 0.010 | (0.100) |
| Visual impairment | 0.017 | (0.129) | 0.010 | (0.101) | 0.006 | (0.078) | 0.006 | (0.075) |
| Cognitive impairment | 0.038 | (0.191) | 0.032 | (0.175) | 0.013 | (0.115) | 0.013 | (0.115) |
| Mobility impairment | 0.067 | (0.250) | 0.037 | (0.189) | 0.014 | (0.118) | 0.012 | (0.107) |
| Difficulty dressing or bathing | 0.020 | (0.140) | 0.012 | (0.108) | 0.002 | (0.048) | 0.002 | (0.045) |
| Difficulty going outside alone | 0.043 | (0.202) | 0.026 | (0.159) | 0.006 | (0.077) | 0.006 | (0.074) |
| **Number of disabilities** | | | | | | | | |
| Zero | 0.880 | (0.325) | 0.924 | (0.265) | 0.958 | (0.200) | 0.963 | (0.188) |
| One | 0.065 | (0.246) | 0.044 | (0.205) | 0.032 | (0.176) | 0.028 | (0.166) |
| Two | 0.027 | (0.163) | 0.018 | (0.131) | 0.007 | (0.081) | 0.006 | (0.076) |
| Three or more | 0.028 | (0.165) | 0.015 | (0.121) | 0.003 | (0.054) | 0.003 | (0.050) |
| **Unemployment rate** | 6.349 | (2.791) | 6.361 | (2.796) | 6.331 | (2.780) | 6.332 | (2.781) |
| **Female** | 0.516 | (0.500) | 0.507 | (0.500) | 0.470 | (0.499) | 0.472 | (0.499) |
| **Race/ethnicity** | | | | | | | | |
| White non-Hispanic | 0.623 | (0.485) | 0.587 | (0.492) | 0.609 | (0.488) | 0.598 | (0.490) |
| Black non-Hispanic | 0.120 | (0.325) | 0.128 | (0.334) | 0.119 | (0.324) | 0.121 | (0.326) |
| Hispanic/Latinx | 0.169 | (0.375) | 0.192 | (0.394) | 0.182 | (0.386) | 0.189 | (0.391) |
| Other race/ethnicity | 0.087 | (0.282) | 0.094 | (0.292) | 0.090 | (0.286) | 0.092 | (0.288) |
| **Age** | | | | | | | | |
| Age 18-34 | 0.290 | (0.454) | 0.372 | (0.483) | 0.343 | (0.475) | 0.369 | (0.482) |
| Age 35-49 | 0.242 | (0.429) | 0.311 | (0.463) | 0.315 | (0.464) | 0.338 | (0.473) |
| Age 49-64 | 0.247 | (0.431) | 0.317 | (0.465) | 0.273 | (0.445) | 0.293 | (0.455) |
| Age 65-99 | 0.220 | (0.415) | -- | 0.000 | 0.069 | (0.254) | -- | 0.000 |
| **Occupation** | | | | | | | | |
| Management | -- | | -- | | 0.179 | (0.384) | 0.177 | (0.381) |
| Professional | -- | | -- | | 0.240 | (0.427) | 0.240 | (0.427) |
| Service | -- | | -- | | 0.161 | (0.367) | 0.162 | (0.368) |
| Sales | -- | | -- | | 0.094 | (0.292) | 0.093 | (0.290) |
| Office, admin. support | -- | | -- | | 0.105 | (0.307) | 0.105 | (0.306) |
| Farming, fishing, and forestry | -- | | -- | | 0.007 | (0.083) | 0.007 | (0.082) |
| Construction, extraction | -- | | -- | | 0.054 | (0.227) | 0.056 | (0.229) |
| Installation, maintenance, repair | -- | | -- | | 0.031 | (0.174) | 0.032 | (0.175) |
| Production | -- | | -- | | 0.053 | (0.223) | 0.054 | (0.225) |
| Transportation | -- | | -- | | 0.076 | (0.265) | 0.076 | (0.265) |
| Armed Forces | -- | | -- | | 0.000 | (0.012) | 0.000 | (0.012) |
| **Observations** | 1,976,071 | | 1,470,035 | | 1,221,664 | | 1,119,402 | |

Source: Authors' computations using Current Population Survey data for May 2020 to April 2022.

**Table A-6: Multinomial logits for results in Table 4**

Figures represent relative risk ratios, representing change in likelihood of this outcome relative to base outcome of no employment

| Employment outcome (base=no employment): | Telework employment (1) | | | Non-telework employment (2) | | | Telework employment (3) | | | Non-telework employment (4) | | |
|---|---|---|---|---|---|---|---|---|---|---|---|---|
| **State unemployment rate** | | | | | | | 0.9424 | (0.0035) | ** | 0.9543 | (0.0025) | ** |
| * No disability | 0.9242 | (0.0067) | ** | 0.9764 | (0.0041) | ** | -- | | | -- | | |
| * Disability | 0.9423 | (0.0035) | ** | 0.9541 | (0.0024) | ** | -- | | | -- | | |
| * Hearing impairment | -- | | | -- | | | 1.0236 | (0.0106) | * | 1.0214 | (0.0062) | ** |
| * Visual impairment | -- | | | -- | | | 0.9356 | (0.0178) | ** | 1.0047 | (0.0100) | |
| * Cognitive impairment | -- | | | -- | | | 0.9541 | (0.0119) | ** | 0.9969 | (0.0075) | |
| * Mobility impairment | -- | | | -- | | | 0.9772 | (0.0109) | * | 1.0170 | (0.0061) | ** |
| * Difficulty dressing or bathing | -- | | | -- | | | 0.9049 | (0.0297) | ** | 1.0054 | (0.0153) | |
| * Difficulty going outside alone | -- | | | -- | | | 0.9748 | (0.0182) | | 0.9924 | (0.0096) | |
| **Female** | 0.7182 | (0.0044) | ** | 0.5557 | (0.0023) | ** | 0.7249 | (0.0045) | ** | 0.5611 | (0.0023) | ** |
| **Race/ethnicity** | | | | | | | | | | | | |
| White non-Hispanic (excl.) | | | | | | | | | | | | |
| Black non-Hispanic | 0.5649 | (0.0066) | ** | 0.7796 | (0.0055) | ** | 0.5719 | (0.0067) | ** | 0.7903 | (0.0056) | ** |
| Hispanic/Latinx | 0.3996 | (0.0043) | ** | 0.9268 | (0.0057) | ** | 0.4011 | (0.0043) | ** | 0.9304 | (0.0057) | ** |
| Other race/ethnicity | 1.0234 | (0.0106) | * | 0.7209 | (0.0057) | ** | 1.0287 | (0.0106) | ** | 0.7247 | (0.0058) | ** |
| **Age** | | | | | | | | | | | | |
| Age 18-34 (excluded) | | | | | | | | | | | | |
| Age 35-49 dummy | 2.1272 | (0.0173) | ** | 1.6209 | (0.0092) | ** | 2.1413 | (0.0174) | ** | 1.6322 | (0.0093) | ** |
| Age 49-64 dummy | 1.0635 | (0.0087) | ** | 1.0053 | (0.0054) | | 1.0674 | (0.0087) | ** | 1.0092 | (0.0054) | |
| Age64-99 dummy | 0.1030 | (0.0012) | ** | 0.1205 | (0.0007) | ** | 0.1009 | (0.0012) | ** | 0.1173 | (0.0007) | ** |
| **State dummies** | Yes | | | Yes | | | Yes | | | Yes | | |
| **State dummies * disability** | Yes | | | Yes | | | No | | | No | | |
| **State dummies * disability types** | No | | | No | | | Yes | | | Yes | | |
| **Month dummies** | Yes | | | Yes | | | Yes | | | Yes | | |
| **Observations** | | | | 1,976,071 | | | | | | 1,976,071 | | |

Source: Authors' computations using Current Population Survey data for May 2020 to April 2022.

* p<.05  ** p<.01
Descriptive statistics in Table A-5

**Table A-7. Additional Tests on Effect of Tight Labor Markets on Telework during the Pandemic**

Marginal effects from multinomial logits showing probability effect of 1-point decline in state unemployment rate. Sample includes employed and non-employed.

|  | All adults age 18+ | | | Working age (18-64) | | |
|---|---|---|---|---|---|---|
|  | Probability effect (1) | (s.e.) (2) | % of addl. jobs in telework (3) | Probability effect (4) | (t-stat) (5) | % of addl. jobs in telework (6) |
| **Panel A: With basic controls^** | | | | | | |
|   No disability | | | | | | |
|     Any employment | 0.0091 | (0.0005) ** | | 0.0093 | (0.0005) ** | |
|     Telework employment | 0.0028 | (0.0003) ** | 30.5% | 0.0031 | (0.0004) ** | 33.8% |
|     Non-telework employment | 0.0063 | (0.0005) ** | | 0.0062 | (0.0006) ** | |
|   Disability | | | | | | |
|     Any employment | 0.0059 | (0.0007) ** | | 0.0065 | (0.0009) ** | |
|     Telework employment | 0.0031 | (0.0003) ** | 51.8% | 0.0034 | (0.0004) ** | 51.2% |
|     Non-telework employment | 0.0028 | (0.0007) ** | | 0.0032 | (0.0009) ** | |
|   n | 1,976,071 | | | 1,470,035 | | |
| **Panel B: Also controlling for major occupation** | | | | | | |
|   No disability | | | | | | |
|     Any employment | 0.0050 | (0.0003) ** | | 0.0051 | (0.0003) ** | |
|     Telework employment | 0.0025 | (0.0004) ** | 50.0% | 0.0027 | (0.0005) ** | 53.3% |
|     Non-telework employment | 0.0025 | (0.0005) ** | | 0.0024 | (0.0005) ** | |
|   Disability | | | | | | |
|     Any employment | 0.0034 | (0.0008) ** | | 0.0037 | (0.0009) ** | |
|     Telework employment | 0.0079 | (0.0009) ** | 229.1% | 0.0072 | (0.0010) ** | 191.5% |
|     Non-telework employment | -0.0044 | (0.0011) ** | | -0.0034 | (0.0013) ** | |
|   n | 1,221,664 | | | 1,119,402 | | |

Source: Authors' computations using Current Population Survey data for May 2020 to April 2022.

* p<.05  ** p<.01

^ Basic controls include dummies for gender, age, race/ethnicity, month, state, and disability*state. Based on multinomial logit results in Table A-8 with descriptive statistics in Table A-5.

**Table A-8: Multinomial logits for results in Table A-7**

Figures represent relative risk ratios, representing change in likelihood of this outcome relative to base outcome of no employment

| Employment outcome (base=no employment): | Age 18+ | | | | Age 18-64 | | | |
|---|---|---|---|---|---|---|---|---|
| | Telework employment (1) | | Non-telework employment (2) | | Telework employment (3) | | Non-telework employment (4) | |
| **Panel A: With basic controls** | | | | | | | | |
| **State unemployment rate** | | | | | | | | |
|   * Disability | 0.9423 (0.0035) | ** | 0.9541 (0.0024) | ** | 0.9411 (0.0037) | ** | 0.9533 (0.0027) | ** |
|   * No disability | 0.9242 (0.0067) | ** | 0.9764 (0.0041) | ** | 0.9285 (0.0074) | ** | 0.9774 (0.0046) | ** |
| **Female** | 0.7182 (0.0044) | ** | 0.5557 (0.0023) | ** | 0.7187 (0.0048) | ** | 0.5526 (0.0025) | ** |
| **Race/ethnicity** | | | | | | | | |
|   White non-Hispanic (excl.) | | | | | | | | |
|   Black non-Hispanic | 0.5649 (0.0066) | ** | 0.7796 (0.0055) | ** | 0.5412 (0.0066) | ** | 0.7442 (0.0056) | ** |
|   Hispanic/Latinx | 0.3996 (0.0043) | ** | 0.9268 (0.0057) | ** | 0.3892 (0.0043) | ** | 0.8990 (0.0058) | ** |
|   Other race/ethnicity | 1.0234 (0.0106) | * | 0.7209 (0.0057) | ** | 1.0098 (0.0108) | | 0.6918 (0.0058) | ** |
| **Age** | | | | | | | | |
|   Age 18-34 (excl.) | | | | | | | | |
|   Age 35-49 | 2.1272 (0.0173) | ** | 1.6209 (0.0092) | ** | 2.1339 (0.0174) | ** | 1.6272 (0.0093) | ** |
|   Age 49-64 | 1.0635 (0.0087) | ** | 1.0053 (0.0054) | | 1.0681 (0.0088) | ** | 1.0090 (0.0054) | |
|   Age 65-99 | 0.1030 (0.0012) | ** | 0.1205 (0.0007) | ** | -- | | -- | |
| **State dummies** | Yes | | Yes | | Yes | | Yes | |
| **State dummies * disability** | Yes | | Yes | | Yes | | Yes | |
| **Month dummies** | Yes | | Yes | | Yes | | Yes | |
| **n** | 1,976,071 | | | | 1,470,035 | | | |
| **Panel B: Also controlling for major occupation** | | | | | | | | |
| **State unemployment rate** | | | | | | | | |
|   * Disability | 0.9090 (0.0051) | ** | 0.9228 (0.0043) | ** | 0.9047 (0.0053) | ** | 0.9201 (0.0045) | ** |
|   * No disability | 0.9169 (0.0094) | ** | 0.9787 (0.0076) | ** | 0.9201 (0.0104) | ** | 0.9756 (0.0084) | ** |

| | | | | | | | | | | | | |
|---|---|---|---|---|---|---|---|---|---|---|---|---|
| **Female** | 0.8423 | (0.0092) | ** | 0.8710 | (0.0083) | ** | 0.8412 | (0.0096) | ** | 0.8747 | (0.0088) | ** |
| **Race/ethnicity** | | | | | | | | | | | | |
|   White non-Hispanic (excl.) | | | | | | | | | | | | |
|   Black non-Hispanic | 0.5676 | (0.0097) | ** | 0.5918 | (0.0081) | ** | 0.5540 | (0.0098) | ** | 0.5818 | (0.0083) | ** |
|   Hispanic/Latinx | 0.6605 | (0.0105) | ** | 0.9026 | (0.0113) | ** | 0.6581 | (0.0108) | ** | 0.9018 | (0.0117) | ** |
|   Other race/ethnicity | 1.0732 | (0.0191) | ** | 0.7236 | (0.0116) | ** | 1.0639 | (0.0195) | ** | 0.7077 | (0.0118) | ** |
| **Age** | | | | | | | | | | | | |
|   Age 18-34 (excl.) | | | | | | | | | | | | |
|   Age 35-49 | 1.6647 | (0.0211) | ** | 1.4893 | (0.0162) | ** | 1.6592 | (0.0211) | ** | 1.4871 | (0.0162) | ** |
|   Age 49-64 | 1.3860 | (0.0180) | ** | 1.4990 | (0.0165) | ** | 1.3823 | (0.0181) | ** | 1.4992 | (0.0165) | ** |
|   Age 65-99 | 0.7452 | (0.0144) | ** | 1.0766 | (0.0167) | ** | -- | | | -- | | |
| **Occupation** | | | | | | | | | | | | |
|   Management (excl.) | | | | | | | | | | | | |
|   Professional | 0.8306 | (0.0145) | ** | 0.9840 | (0.0166) | | 0.8415 | (0.0158) | ** | 1.0338 | (0.0188) | |
|   Service | 0.0364 | (0.0008) | ** | 0.5732 | (0.0091) | ** | 0.0349 | (0.0008) | ** | 0.5644 | (0.0096) | ** |
|   Sales | 0.2157 | (0.0045) | ** | 0.6910 | (0.0129) | ** | 0.2063 | (0.0046) | ** | 0.6789 | (0.0135) | ** |
|   Office, admin. support | 0.3696 | (0.0075) | ** | 0.7340 | (0.0138) | ** | 0.3645 | (0.0078) | ** | 0.7397 | (0.0148) | ** |
|   Farming, fishing, and forestry | 0.0180 | (0.0020) | ** | 0.5422 | (0.0239) | ** | 0.0171 | (0.0019) | ** | 0.5260 | (0.0245) | ** |
|   Construction, extraction | 0.0257 | (0.0010) | ** | 0.5819 | (0.0125) | ** | 0.0241 | (0.0010) | ** | 0.5762 | (0.0130) | ** |
|   Installation, maintenance, repair | 0.0717 | (0.0031) | ** | 1.0045 | (0.0300) | | 0.0697 | (0.0032) | ** | 1.0124 | (0.0318) | |
|   Production | 0.0477 | (0.0017) | ** | 0.7461 | (0.0166) | ** | 0.0462 | (0.0017) | ** | 0.7506 | (0.0176) | ** |
|   Transportation | 0.0215 | (0.0008) | ** | 0.5839 | (0.0111) | ** | 0.0211 | (0.0008) | ** | 0.5864 | (0.0118) | ** |
|   Armed Forces | 0.0000 | (0.0000) | ** | 0.0000 | (0.0000) | ** | 0.0000 | (0.0000) | ** | 0.0000 | (0.0000) | ** |
| **State dummies** | Yes | | | Yes | | | Yes | | | Yes | | |
| **State dummies * disability** | Yes | | | Yes | | | Yes | | | Yes | | |
| **Month dummies** | Yes | | | Yes | | | Yes | | | Yes | | |
| **N** | | 1,221,664 | | | | | | 1,119,402 | | | | |

Source: Authors' computations using Current Population Survey data for May 2020 to April 2022.
* p<.05  ** p<.01
See Table A-5 for descriptive statistics

**Table A-9: Effect of tight labor markets on telework during pandemic by number of disabilities**

Marginal effects from multinomial logits showing probability effect of 1-point decline in state unemployment rate, for all adults age 18+

|  | Probability effect (1) | (t-stat) (2) |  | % of addl. jobs in telework (3) |
|---|---|---|---|---|
| No disability |  |  |  |  |
|    Any employment | 0.0091 | (0.0005) | ** |  |
|    Telework employment | 0.0028 | (0.0003) | ** | 30.3% |
|    Non-telework employment | 0.0064 | (0.0005) | ** |  |
| One disability condition |  |  |  |  |
|    Any employment | 0.0062 | (0.0009) | ** |  |
|    Telework employment | 0.0036 | (0.0004) | ** | 58.8% |
|    Non-telework employment | 0.0026 | (0.0009) | ** |  |
| Two disability conditions |  |  |  |  |
|    Any employment | 0.0070 | (0.0011) | ** |  |
|    Telework employment | 0.0025 | (0.0005) | ** | 35.3% |
|    Non-telework employment | 0.0045 | (0.0010) | ** |  |
| Three or more disability conditions |  |  |  |  |
|    Any employment | 0.0055 | (0.0010) | ** |  |
|    Telework employment | 0.0024 | (0.0005) | ** | 43.9% |
|    Non-telework employment | 0.0031 | (0.0009) | ** |  |
|  |  |  |  |  |
| n | 1,976,071 |  |  |  |

Source: Authors' computations using Current Population Survey data for May 2020 to April 2022.
* p<.05  ** p<.01

Controls include dummies for gender, age, race/ethnicity, month, state, and disability*state. Based on multinomial logit results in Table A-10 with descriptive statistics in Table A-5.

**Table A-10: Multinomial logits for results in Table A-9**

Figures represent relative risk ratios, representing change in likelihood of this outcome relative to base outcome of no employment

| Employment outcome (base=no employment): | Telework employment (1) | | | Non-telework employment (2) | | |
|---|---|---|---|---|---|---|
| **State unemployment rate** | 0.941954 | -0.00348 | ** | 0.9537 | (0.0025) | ** |
| * One disability | 0.98613 | -0.00696 | * | 1.0254 | (0.0042) | ** |
| * Two disabilities | 0.962344 | -0.01652 | * | 1.0050 | (0.0085) | |
| * Three or more disabilities | 0.910087 | -0.02707 | ** | 0.9990 | (0.0127) | |
| | | | | | | |
| **Female** | 0.71828 | -0.00445 | ** | 0.5558 | (0.0023) | ** |
| **Race/ethnicity** | | | | | | |
| White non-Hispanic (excl.) | | | | | | |
| Black non-Hispanic | 0.567981 | -0.00665 | ** | 0.7842 | (0.0056) | ** |
| Hispanic/Latinx | 0.40019 | -0.00433 | ** | 0.9282 | (0.0057) | ** |
| Other race/ethnicity | 1.025356 | -0.01059 | * | 0.7223 | (0.0058) | ** |
| **Age** | | | | | | |
| Age 18-34 (excluded) | | | | | | |
| Age 35-49 dummy | 2.140392 | -0.01743 | ** | 1.6316 | (0.0093) | ** |
| Age 49-64 dummy | 1.070339 | -0.00875 | ** | 1.0125 | (0.0054) | * |
| Age64-99 dummy | 0.103061 | -0.00121 | ** | 0.1206 | (0.0007) | ** |
| | | | | | | |
| **State dummies** | Yes | | | Yes | | |
| **State dummies * disability number** | Yes | | | Yes | | |
| **Month dummies** | Yes | | | Yes | | |
| | | | | | | |
| **Observations** | | | 1,976,071 | | | |

Source: Authors' computations using Current Population Survey data for May 2020 to April 2022.

* p<.05  ** p<.01
See Table A-5 for descriptive statistics